\pdfoutput=1

\documentclass[11pt, 
final,
onecolumn,
]{article}

\usepackage[english]{babel}
\usepackage[T1]{fontenc}
\usepackage[utf8]{inputenc}
\usepackage{microtype}
\usepackage{lmodern}
\usepackage{csquotes}

\usepackage{amsmath, amsfonts, amssymb}
\usepackage{mathabx}

\usepackage[totalwidth=480pt, totalheight=680pt]{geometry}

\usepackage[affil-it]{authblk}

\usepackage{booktabs}
	
\newcommand{\LESAvg}[1]{ \widebar{#1} }
\newcommand{\tr}[1]{\operatorname{tr}(#1)}

\usepackage[
	backend=bibtex8,
	sortcites=true,
	sorting=none,
	style=numeric,
]{biblatex}
\DeclareNameAlias{default}{last-first}
\AtEveryBibitem{\ifentrytype{article}{\clearfield{title}}{}}
\AtEveryBibitem{\ifentrytype{inproceedings}{\clearfield{title}}{}}
\AtEveryBibitem{\ifentrytype{incollection}{\clearfield{title}}{}}

\ifpdf
	\usepackage{hyperref}
	\hypersetup{
		bookmarks=true,
		bookmarksnumbered=true,
		bookmarksopen=true,
		colorlinks=true,
		allcolors=blue,
		pdfborder=0 0 0,
		pdftitle={Physically-consistent subgrid-scale models for large-eddy simulation of incompressible turbulent flows},pdfauthor={Maurits H. Silvis, Roel Verstappen}
	}
\fi

\usepackage[capitalize, sort]{cleveref}

\title{\vspace{-0.6\baselineskip}Physically-consistent subgrid-scale models for large-eddy simulation of incompressible turbulent flows}
\author{Maurits H. Silvis\thanks{Email address: \href{mailto:m.h.silvis@rug.nl}{m.h.silvis@rug.nl}} }
\author{Roel Verstappen}
\affil{Johann Bernoulli Institute for Mathematics and Computer Science, University of Groningen, The Netherlands\vspace{-0.8\baselineskip}}
\begin{document}

\maketitle

\vspace{-1.9\baselineskip}

\paragraph{Abstract} Assuming a general constitutive relation for the turbulent stresses in terms of the local large-scale velocity gradient, we constructed a class of subgrid-scale models for large-eddy simulation that are consistent with important physical and mathematical properties.
In particular, they preserve symmetries of the Navier-Stokes equations and exhibit the proper near-wall scaling.
They furthermore show desirable dissipation behavior and are capable of describing nondissipative effects.
We provided examples of such physically-consistent models and showed that existing subgrid-scale models do not all satisfy the desired properties.

\section{Introduction}
\label{sec:intro}

It is well known that the governing equations of fluid dynamics, the Navier-Stokes equations, are form invariant under certain transformations, such as instantaneous rotations of the coordinate system and the Galilean transformation~\cite{pope11}.
These transformations, also referred to as symmetries of the equations, play an important physical role, because they ensure that the description of fluids is the same in all inertial frames of reference. 
Furthermore, they relate to conservation and scaling laws~\cite{razafindralandyetal07}.
It has since long been realized that it is desirable that the basic equations of large-eddy simulation, which are used to obtain information about the large-scale behavior of turbulent flows, also satisfy these physical principles. 
Speziale~\cite{speziale85} was the first to emphasize the importance of Galilean invariance of subgrid-scale models for large-eddy simulation. 
Later, Oberlack~\cite{oberlack97} formulated requirements to make subgrid-scale models compatible with all the symmetries of the Navier-Stokes equations.
Besides preserving the symmetries of the underlying equations, it is desirable that subgrid-scale models have the same basic properties as the turbulent stresses, such as the observed near-wall scaling~\cite{chapmankuhn86} and the dissipation behavior~\cite{vreman04}. 

In the current work we aim to construct subgrid-scale models that preserve these and other exact properties of the Navier-Stokes equations and the turbulent stresses. 
As starting point, we assume a general class of models based on the local large-scale velocity gradient~\cite{pope75,lundnovikov92}.
Aside from offering a framework for subgrid-scale models, it allows us to go beyond the purely dissipative description of turbulent flows that is typically provided by eddy viscosity models.
We will look to set the model coefficients in such a way that the desired properties are satisfied.

The outline of this paper is as follows. The class of nonlinear subgrid-scale models of~\cite{pope75,lundnovikov92} is introduced in \cref{sec:GNM}, where it is also reduced to a set
of independent terms.
In \cref{sec:modconstraints} we discuss several properties of the Navier-Stokes equations and the turbulent stresses, as well as the resulting requirements for the form of subgrid-scale models.
In particular, we obtain constraints on the coefficients of the class of nonlinear models of \cref{sec:GNM}.
An analysis of properties of existing subgrid-scale models is performed in \cref{sec:modanalysis}, after which, in \cref{sec:modexamples}, we provide examples of physically-consistent models that arise when said requirements are applied to constrain the class of nonlinear models.
Finally, \cref{sec:concldisc} consists of concluding remarks and suggestions for future work.

\section{General nonlinear subgrid-scale models}
\label{sec:GNM}

Large-eddy simulation is aimed at predicting the large-scale behavior of turbulent flows. 
To distinguish between large and small scales of motion, usually a filtering or coarse-graining operation is used, indicated by an overbar, here. 
The evolution of the large-scale velocity field is governed by the filtered Navier-Stokes equations,~\cite{sagaut06}
\begin{align}
\label{eq:filtns}
\begin{alignedat}{2}
\frac{ \partial \LESAvg{ u }_i }{ \partial x_i }  & = 0, \qquad\qquad & 
\frac{ \partial \LESAvg{ u }_i }{ \partial t } + \LESAvg{ u }_j \frac{ \partial \LESAvg{ u }_i }{ \partial x_j } & = -\frac{ 1 }{ \rho } \frac{ \partial \LESAvg{ p } }{ \partial x_i } + \nu \frac{ \partial^2 \LESAvg{ u }_i }{ \partial x_j \partial x_j } - \frac{ \partial }{ \partial x_j } \tau_{ij}.
\end{alignedat}
\end{align}
Here, the turbulent or subgrid-scale stresses, $\tau_{ij} = \LESAvg{ u_i u_j } - \LESAvg{ u }_i \LESAvg{ u }_j$, represent the interactions between large and small scales of motion.
Following~\cite{pope75,lundnovikov92}, we will assume they can be modeled using the following constitutive relation, based only on the local filtered velocity gradient,
\begin{equation}
\label{eq:assumption}
\tau_{ij}^{mod} = f(\LESAvg{S}_{kl}, \LESAvg{\Omega}_{kl}),
\end{equation}
where the filtered rate-of-strain and rate-of-rotation tensors are given by
\begin{align}
\label{eq:SandW}
\begin{alignedat}{2}
\LESAvg{S}_{ij} & = \frac{ 1 }{ 2 }\left( \frac{ \partial \LESAvg{ u }_i }{ \partial x_j } + \frac{ \partial \LESAvg{ u }_j }{ \partial x_i } \right), \qquad\qquad &
\LESAvg{\Omega}_{ij} & = \frac{ 1 }{ 2 }\left( \frac{ \partial \LESAvg{ u }_i }{ \partial x_j } - \frac{ \partial \LESAvg{ u }_j }{ \partial x_i } \right).
\end{alignedat}
\end{align}
In anticipation of \cref{sec:modconstraints}, we will further assume that the relation of \cref{eq:assumption} is isotropic, i.e., independent of the angle of observation. 
As a consequence, the subgrid-scale stress model can consist of all terms of the form $\LESAvg{S}^{m_1}\LESAvg{\Omega}^{n_1}\LESAvg{S}^{m_2}\LESAvg{\Omega}^{n_2}\ldots$, where $n_i$, $m_i$ are positive integers.
The Cayley-Hamilton theorem of matrix algebra allows for a reduction to a finite number of model terms~\cite{rivlin55,spencerrivlin58,spencerrivlin62}. Symmetrizing the result, we obtain~\cite{pope75,lundnovikov92}
\begin{align}
\label{eq:tensors}
\begin{alignedat}{3}
T_0 & = I,                 \qquad\qquad & T_4 & = \LESAvg{S}\LESAvg{\Omega} - \LESAvg{\Omega}\LESAvg{S},                                   \qquad\qquad & T_8 & = \LESAvg{S}\LESAvg{\Omega}\LESAvg{S}^2 - \LESAvg{S}^2\LESAvg{\Omega}\LESAvg{S}, \\
T_1 & = \LESAvg{S},        \qquad\qquad & T_5 & = \LESAvg{S}^2\LESAvg{\Omega} - \LESAvg{\Omega}\LESAvg{S}^2,                               \qquad\qquad & T_9 & = \LESAvg{S}^2\LESAvg{\Omega}^2 + \LESAvg{\Omega}^2\LESAvg{S}^2, \\
T_2 & = \LESAvg{S}^2,      \qquad\qquad & T_6 & = \LESAvg{S}\LESAvg{\Omega}^2 + \LESAvg{\Omega}^2\LESAvg{S},                               \qquad\qquad & T_{10} & = \LESAvg{\Omega}\LESAvg{S}^2\LESAvg{\Omega}^2 - \LESAvg{\Omega}^2\LESAvg{S}^2\LESAvg{\Omega}.\\
T_3 & = \LESAvg{\Omega}^2, \qquad\qquad & T_7 & = \LESAvg{\Omega}\LESAvg{S}\LESAvg{\Omega}^2 - \LESAvg{\Omega}^2\LESAvg{S}\LESAvg{\Omega}, \qquad\qquad & ~ & ~
\end{alignedat}
\end{align}
A general class of subgrid-scale models arises from a sum of the above terms, each multiplied by a coefficient that, by the assumptions of \cref{eq:assumption} and isotropy, can depend only on the following tensor invariants~\cite{spencerrivlin62},
\begin{align}
\label{eq:tensorinvariants}
\begin{alignedat}{3}
I_1 & = \tr{\LESAvg{S}^2},      \qquad\qquad & I_3 & = \tr{\LESAvg{S}^3},                \qquad\qquad & I_5 & = \tr{\LESAvg{S}^2\LESAvg{\Omega}^2},\\
I_2 & = \tr{\LESAvg{\Omega}^2}, \qquad\qquad & I_4 & = \tr{\LESAvg{S}\LESAvg{\Omega}^2}, \qquad\qquad & I_6 & = \tr{\LESAvg{S}^2\LESAvg{\Omega}^2\LESAvg{S}\LESAvg{\Omega}}.
\end{alignedat}
\end{align}
In practical tests, usually a smaller number of model terms is used, cf.~for instance~\cite{wangbergstrom05,marstorpetal09}. For an extensive review of the use of these and similar nonlinear models in the RANS community, refer to~\cite{gatskijongen00}.
Rather than discarding any model terms on beforehand, we propose to systematically find the independent contributions using the Gram-Schmidt orthogonalization procedure.
Denoting orthogonalized tensors with a prime, we find
\begin{align}
\label{eq:tensorsorth}
\begin{split}
\begin{alignedat}{2}
T_0' & = I,                                                                   \qquad & T_3' & = \LESAvg{\Omega}^2 - \frac{ I_2 }{ 3 } I - \frac{ I_4 }{ I_1 } \LESAvg{S} - \frac{ I_5 - \frac{ I_1 I_2 }{ 3 } - \frac{ I_3 I_4 }{ I_1 } }{ \frac{ I_1^2 }{ 6 } - \frac{ I_3^2 }{ I_1 } } T_2', \\
T_1' & = \LESAvg{S},                                                          \qquad & T_5' & = \LESAvg{S}^2 \LESAvg{\Omega} - \LESAvg{\Omega} \LESAvg{S}^2 - \frac{ - I_1 I_4 - I_2 I_3 }{ I_1 I_2 - 6 I_5} ( \LESAvg{S} \LESAvg{\Omega} - \LESAvg{\Omega} \LESAvg{S} ),\\
T_2' & = \LESAvg{S}^2 - \frac{ I_1 }{ 3 } I - \frac{ I_3 }{ I_1 } \LESAvg{S}, \qquad & T_6' & = \LESAvg{S} \LESAvg{\Omega}^2 + \LESAvg{\Omega}^2 \LESAvg{S} - \frac{ 2 I_4 }{ 3 } I - \frac{ 2 I_5 }{ I_1 } \LESAvg{S} - \frac{ \frac{ I_2 I_4 }{ 3 } - \frac{ 2 I_4 I_5 }{ I_1 } }{ \frac{ I_2^2 }{ 6 } - \frac{ I_4^2 }{ I_1} } T_3', \\
T_4' & = \LESAvg{S} \LESAvg{\Omega} - \LESAvg{\Omega} \LESAvg{S},             \qquad & T_7' & = \LESAvg{\Omega}\LESAvg{S}\LESAvg{\Omega}^2 - \LESAvg{\Omega}^2\LESAvg{S}\LESAvg{\Omega} - \frac{ -I_1 I_2^2 + 4 I_2 I_5 + 2 I_4^2 }{ I_1 I_2 - 6 I_5 } ( \LESAvg{S} \LESAvg{\Omega} - \LESAvg{\Omega} \LESAvg{S} ).
\end{alignedat}
\end{split}
\end{align}
Note that the expressions for $T_6'$ and $T_7'$ are valid only in case of an axisymmetric strain, i.e., when $T_2'= T_5' = 0$. Otherwise they have to be replaced by zero. The orthogonalized tensors $T_8'$ to $T_{10}'$ are always zero.
\cref{eq:assumption} and the condition of isotropy thus provide a general class of nonlinear models,
\begin{align}
\label{eq:generalnonlinearmodel}
\begin{alignedat}{2}
\tau^{mod} & = \sum_{i = 0}^7 \alpha_i' T_i', \qquad\qquad & \alpha_i' & = \alpha_i'(I_1, I_2, \ldots, I_6).
\end{alignedat}
\end{align}
Its main advantage over formulations based on the original tensors, \cref{eq:tensors}, is that it consists of independent terms that, thus, represent distinct physical processes.
For instance, only $T_1'$ relates to dissipation.
The other terms allow for a description of nondissipative processes.
\cref{eq:generalnonlinearmodel} represents a class of models, because the model coefficients are isotropic but otherwise undetermined. 
In the next section we will therefore discuss several model requirements and determine how these can be satisfied by setting the explicit dependence of the model coefficients on the tensor invariants of \cref{eq:tensorinvariants}.

\section{Model constraints}
\label{sec:modconstraints}

As was alluded to in \cref{sec:intro}, the Navier-Stokes equations and the subgrid-scale stresses have several interesting physical and mathematical properties. 
To make sure that related special properties of solutions are preserved, it is desirable that the equations of large-eddy simulation, \cref{eq:filtns} with a subgrid-scale model replacing the turbulent stresses, satisfy the same principles.
In \cref{sec:modconstraintssymm} we will therefore consider the invariance properties of the Navier-Stokes equations, whereas \cref{sec:modconstraintsnearwallscal} discusses the near-wall scaling behavior of the subgrid-scale stresses. Constraints on the production of subgrid-scale kinetic energy in turbulent flows are the topic of \cref{sec:modconstraintsSGSTKE}.
They lead to several requirements on the form of subgrid-scale models in general, that are subsequently applied to the class of nonlinear models of the previous section. 

\subsection{Symmetry requirements}
\label{sec:modconstraintssymm}

The unfiltered Navier-Stokes equations are invariant under the following transformations~\cite{pope11,oberlack97,oberlack02,razafindralandyetal07}:
\begin{align}
\label{eq:nssymmtimetrans}
\mbox{S1: } (t, x_i, u_i, p, \nu) & \rightarrow (t + T, x_i, u_i, p, \nu), \\
\label{eq:nssymmprestrans}
\mbox{S2: } (t, x_i, u_i, p, \nu) & \rightarrow (t, x_i, u_i, p + P(t), \nu), \\
\label{eq:nssymmgenGaltrans}
\mbox{S3: } (t, x_i, u_i, p, \nu) & \rightarrow (t, x_i + X_i(t), u_i + \dot{X}_i(t), p - \rho x_i \ddot{X}_i(t), \nu), \\
\label{eq:nssymmorthtrans}
\mbox{S4: } (t, x_i, u_i, p, \nu) & \rightarrow (t, Q_{ij} x_j, Q_{ij} u_j, p, \nu), \\
\label{eq:nssymmscaltrans}
\mbox{S5: } (t, x_i, u_i, p, \nu) & \rightarrow (e^{2a} t, e^{a + b} x_i, e^{-a + b} u_i, e^{-2a + 2b} p, e^{2b} \nu), \\
\label{eq:nssymm2DMFI}
\mbox{S6: } (t, x_i, u_i, p, \nu) & \rightarrow (t, R_{ij}(t) x_j, R_{ij}(t)u_j + \dot{R}_{ij}(t)x_j, p + \tfrac{1}{2} \rho \omega_3^2 (x_1^2 + x_2^2) + 2 \rho \omega_3 \psi, \nu).
\end{align}
In the limit of an inviscid flow, $\nu \rightarrow 0$, the equations allow for an additional symmetry~\cite{oberlack02}:
\begin{align}
\label{eq:eusymmtimerev}
\mbox{S7: } (t, x_i, u_i, p) & \rightarrow (-t, x_i, -u_i, p).
\end{align}
In the time~(\hyperref[eq:nssymmtimetrans]{S1}) and pressure~(\hyperref[eq:nssymmprestrans]{S2}) translations, $T$ and $P(t)$ indicate an arbitrary time shift and a time variation of the (background) pressure, respectively.
The generalized Galilean transformation~(\hyperref[eq:nssymmgenGaltrans]{S3}) encompasses the space translation for $X_i(t)$ constant, and the classical Galilean transformation for $X_i(t)$ linear in time. 
Orthogonal transformations~(\hyperref[eq:nssymmorthtrans]{S4}) are represented by a time-independent matrix $Q$ with $Q_{ik} Q_{jk} = \delta_{ij}$, and describe instantaneous rotations and reflections of the coordinate system.
The scaling transformations~(\hyperref[eq:nssymmscaltrans]{S5}) are parametrized by real $a$ and $b$.
Two-dimensional material frame-indifference (2D MFI, \hyperref[eq:nssymm2DMFI]{S6}) represents a time-dependent but constant-in-rate rotation of the coordinate system about the $x_3$ axis, characterized by a rotation matrix $R(t)$ with $\dot{R}_{ik} R_{jk} = \epsilon_{3ij}\omega_3$, for constant $\omega_3$. Here, the flow is assumed to be confined to the $x_1$ and $x_2$ directions, and $\psi$ represents its two-dimensional stream function.
The final transformation will be referred to as time reversal~(\hyperref[eq:eusymmtimerev]{S7}).

We now require that each of the above symmetry transformations also applies to the filtered Navier-Stokes equations, \cref{eq:filtns}, with a subgrid-scale model in place of the turbulent stresses.
It is assumed that explicit filtering does not destroy symmetry properties.
The requirements on the form of subgrid-scale models that result from this~\cite{oberlack97} are summarized in the left-hand column of \cref{tab:modconstraints}.
Application of these conditions to the class of nonlinear models of \cref{eq:generalnonlinearmodel} leads to several constraints on the form of the model coefficients, as shown in the middle column of the same table. The column on the right may help in selecting the explicit functional dependence of the coefficients on the tensor invariants, \cref{eq:tensorinvariants}, by showing the transformation behavior of several quantities.

With reference to \cref{tab:modconstraints}, we remark that time~(\hyperref[eq:nssymmtimetrans]{S1}), pressure~(\hyperref[eq:nssymmprestrans]{S2}) and generalized Galilean~(\hyperref[eq:nssymmgenGaltrans]{S3}) invariance are automatically satisfied by subgrid-scale models that are based on the filtered velocity gradient alone and, thus, also by the class of nonlinear models of \cref{eq:generalnonlinearmodel}. 
Furthermore, these models satisfy rotation and reflection invariance~(\hyperref[eq:nssymmorthtrans]{S4}) by construction (refer to \cref{sec:GNM}).

Invariance under scaling transformations~(\hyperref[eq:nssymmscaltrans]{S5}), which relates to the appearance of wall and other scaling laws, is not straightforwardly satisfied, because it requires an internal length scale. Neither the velocity gradient, nor the externally-imposed large-eddy simulation filter length, $\LESAvg{\delta}$, can provide this~\cite{oberlack97,razafindralandyetal07}. 
If the dynamic procedure~\cite{germanoetal91} is used to determine the model constants, scale invariance is known to be restored~\cite{oberlack97,razafindralandyetal07}. 
Use of the dynamic procedure is beyond the scope of the current study, but let us note that it relies on a second filtering operation, which may destroy some of the other symmetry properties unless certain severe restrictions on the filter are fulfilled~\cite{razafindralandyetal07}.

A peculiar symmetry of the Navier-Stokes equations is material frame-indifference in the limit of a two-dimensional flow~(2D MFI, \hyperref[eq:nssymm2DMFI]{S6}). Despite its name, this is not a material but a frame invariance property: as explained above, for two-dimensional flows, the Navier-Stokes equations hold in inertial as well as in certain constantly-rotating frames.
Although it seems unlikely that a turbulent flow becomes two-dimensional as a whole, there is no reason this cannot occur locally. Consider the flow near a wall, for instance. Therefore, two-dimensionality is interpreted as a local property throughout this paper. Locally two-dimensional flows can be characterized by the following set of invariants.
\begin{align}
\label{eq:2Dflowchar}
I_3 = I_4 = I_5 - \frac{ 1 }{ 2 } I_1 I_2 = I_6 = 0.
\end{align}
2D MFI provides rather restrictive constraints on the form of subgrid-scale models, although no information can be obtained about model coefficients $\alpha_5'$ to $\alpha_7'$, because the corresponding tensors vanish in locally two-dimensional flows.

As for reversibility~(\hyperref[eq:eusymmtimerev]{S7}) of the turbulent stresses, application of
the dynamic procedure~\cite{germanoetal91} can, in principle, ensure time reversal invariance~\cite{caratietal01}, although, again, care has to be taken that none of the other symmetries are broken by the test filter~\cite{razafindralandyetal07}.

\subsection{Near-wall scaling requirements}
\label{sec:modconstraintsnearwallscal}

Using high-resolution numerical simulations, Chapman and Kuhn~\cite{chapmankuhn86} have revealed the near-wall scaling behavior of the turbulent stresses. They observed that, 
in the vicinity of a wall, the fluctuations of the tangential velocity components show a linear scaling with the distance from that wall (denoted $x_2$ in this paper). 
The incompressibility constraint then leads to a quadratic scaling for the wall-normal velocity fluctuations, so that the scaling behavior of the time-averaged turbulent stresses can be derived.
In what follows, we will require that modeled subgrid-scale stresses show the same near-wall behavior, but then instantaneously.
(Refer to property P1 in \cref{tab:modconstraints}.)
This ensures that, for instance, dissipative effects fall off quickly enough near walls. 
A summary of the constraints on the coefficients of the class of nonlinear models, \cref{eq:generalnonlinearmodel}, and of the behavior of a few selected quantities is provided in \cref{tab:modconstraints}, as well.
Note that this principle provides no information about coefficients $\alpha_6'$ and $\alpha_7'$.

\begin{table}
	\centering
	\caption{
		Summary of model requirements~\cite{oberlack97,oberlack02,chapmankuhn86}, the resulting constraints on the coefficients of the class of nonlinear models, \cref{eq:generalnonlinearmodel}, and the behavior of a few selected quantities that may be used to construct the model coefficients.
		The properties considered are 
		\hyperref[eq:nssymmtimetrans]{S1}, \hyperref[eq:nssymmprestrans]{S2}, \hyperref[eq:nssymmgenGaltrans]{S3}:~time, pressure and generalized Galilean invariance;
		\hyperref[eq:nssymmorthtrans]{S4}:~rotation and reflection invariance;
		\hyperref[eq:nssymmscaltrans]{S5}:~scaling invariance;
		\hyperref[eq:nssymm2DMFI]{S6}:~two-dimensional material frame-indifference;
		\hyperref[eq:eusymmtimerev]{S7}:~time reversal invariance;
		and 
		\hyperref[sec:modconstraintsnearwallscal]{P1}:~near-wall scaling behavior, \cref{sec:modconstraintsnearwallscal}.
		Note that property \hyperref[eq:nssymm2DMFI]{S6} assumes a flow that is locally two-dimensional, \cref{eq:2Dflowchar},
		and that transformed variables are denoted with a hat.
	}
	\label{tab:modconstraints}
	\begin{tabular} {p{0.05\linewidth} p{0.25\linewidth} p{0.29\linewidth} p{0.29\linewidth}}
		\toprule
		     & \textbf{Model requirement}                               & \textbf{Coefficient constraints}                     & \textbf{Selected quantities} \\
		\midrule
		S1-3 & $\hat{\tau}_{ij}^{mod} = \tau_{ij}^{mod}$                & $\hat{\alpha}_i = \alpha_i$                          & $\hat{I}_i = I_i$ \\
		\midrule
		S4   & $\hat{\tau}_{ij}^{mod} = Q_{im} Q_{jn} \tau_{mn}^{mod}$  & $\hat{\alpha}_i = \alpha_i$                          & $\hat{I}_i = I_i$ \\
		\midrule
		S5   & $\hat{\tau}_{ij}^{mod} = e^{-2 a + 2 b} \tau_{ij}^{mod}$ & $\hat{\alpha}_0' = e^{ -2 a + 2 b } \alpha_0'$       &  $\hat{I}_1 = e^{  -4 a } I_1$ \\  
		     &                                                          & $\hat{\alpha}_1' = e^{        2 b } \alpha_1'$       &  $\hat{I}_2 = e^{  -4 a } I_2$ \\
		     &                                                          & $\hat{\alpha}_2' = e^{  2 a + 2 b } \alpha_2'$       &  $\hat{I}_3 = e^{  -6 a } I_3$ \\
		     &                                                          & $\hat{\alpha}_3' = e^{  2 a + 2 b } \alpha_3'$       &  $\hat{I}_4 = e^{  -6 a } I_4$ \\
		     &                                                          & $\hat{\alpha}_4' = e^{  2 a + 2 b } \alpha_4'$       &  $\hat{I}_5 = e^{  -8 a } I_5$ \\
		     &                                                          & $\hat{\alpha}_5' = e^{  4 a + 2 b } \alpha_5'$       &  $\hat{I}_6 = e^{ -12 a } I_6$ \\
		     &                                                          & $\hat{\alpha}_6' = e^{  4 a + 2 b } \alpha_6'$       &  $\hat{\LESAvg{\delta}} = \LESAvg{\delta}$ \\
		     &                                                          & $\hat{\alpha}_7' = e^{  6 a + 2 b } \alpha_7'$       &  \\
		\midrule
		S6   & For 2D flows:                                            & $\hat{\alpha}_0' = \alpha_0'$                        & $\hat{I}_1 = I_1$ \\
		     & $\hat{\tau}_{ij}^{mod} = R_{im} R_{jn} \tau_{mn}^{mod}$  & $\hat{\alpha}_1' = \alpha_1'$                        & $\hat{I}_2 \neq I_2$ \\
		     &                                                          & $\hat{\alpha}_2' = \alpha_2'$                        & $\hat{I}_3 = I_3 = 0$ \\
		     &                                                          & $\alpha_0' = \frac{ 1 }{ 3 } I_1 \alpha_2'$          & $\hat{I}_4 = I_4 = 0$ \\
		     &                                                          & $\hat{\alpha}_3' = \alpha_3' = 0$ when $I_1 = 0$     & $\hat{I}_5 - \frac{ 1 }{ 2 } \hat{I}_1 \hat{I}_2 = I_5 - \frac{ 1 }{ 2 } I_1 I_2 = 0$ \\
		     &                                                          & $\hat{\alpha}_4' = \alpha_4' = 0$ when $I_1 \neq 0$  & $\hat{I}_5 / \hat{I}_2 = I_5 / I_2$ \\
		     &                                                          &                                                      & $\hat{I}_6 = I_6 = 0$ \\                                                       
		\midrule
		S7   & $\hat{\tau}_{ij}^{mod} = \tau_{ij}^{mod}$                & $\hat{\alpha}_0' = \alpha_0'$                        & $\hat{I}_1 = I_1$ \\
		     &                                                          & $\hat{\alpha}_1' = -\alpha_1'$                       & $\hat{I}_2 = I_2$ \\
	      	 &                                                          & $\hat{\alpha}_2' = \alpha_2'$                        & $\hat{I}_3 = -I_3$ \\
		     &                                                          & $\hat{\alpha}_3' = \alpha_3'$                        & $\hat{I}_4 = -I_4$ \\
		     &                                                          & $\hat{\alpha}_4' = \alpha_4'$                        & $\hat{I}_5 = I_5$ \\
		     &                                                          & $\hat{\alpha}_5' = -\alpha_5'$                       & $\hat{I}_6 = I_6$ \\
		     &                                                          & $\hat{\alpha}_6' = -\alpha_6'$                       & \\
		     &                                                          & $\hat{\alpha}_7' = \alpha_7'$                        & \\
		\midrule
		P1   & $\tau_{11}^{mod} = \mathcal{O}(x_2^2)$                   & $\alpha_0' = \mathcal{O}(x_2^4)$                     & $I_1 = \mathcal{O}(1)$ \\
		     & $\tau_{12}^{mod} = \mathcal{O}(x_2^3)$                   & $\alpha_1' = \mathcal{O}(x_2^3)$                     & $I_2 = \mathcal{O}(1)$ \\
		     & $\tau_{13}^{mod} = \mathcal{O}(x_2^2)$                   & $\alpha_2' = \mathcal{O}(x_2^4)$                     & $I_3 = \mathcal{O}(x_2)$ \\
		     & $\tau_{22}^{mod} = \mathcal{O}(x_2^4)$                   & $\alpha_3' = \mathcal{O}(x_2^2)$                     & $I_4 = \mathcal{O}(x_2)$ \\
		     & $\tau_{23}^{mod} = \mathcal{O}(x_2^3)$                   & $\alpha_4' = \mathcal{O}(x_2^4)$                     & $I_5 = \mathcal{O}(1)$ \\
		     & $\tau_{33}^{mod} = \mathcal{O}(x_2^2)$                   & $\alpha_5' = \mathcal{O}(x_2)$                       & $I_6 = \mathcal{O}(x_2^2)$ \\
		     &                                                          &                                                      & $I_1 + I_2 = \mathcal{O}(x_2^2)$ \\
		     &                                                          &                                                      & $I_3 + 3 I_4 = \mathcal{O}(x_2^3)$ \\
		     &                                                          &                                                      & $I_5 - \frac{ 1 }{ 2 } I_1 I_2 = \mathcal{O}(x_2^2)$ \\
		\bottomrule
	\end{tabular}
\end{table}

\subsection{Requirements relating to the production of subgrid-scale kinetic energy}
\label{sec:modconstraintsSGSTKE}

In the current section we will look at energy transport in turbulent flows. In particular, we focus on the transport of energy to small scales of motion, also referred to as production of subgrid-scale kinetic energy or subgrid dissipation. Mathematically it is described by
\begin{align}
\label{eq:subgriddissipation}
D_{\tau} & = -\tr{ \tau \LESAvg{S} }.
\end{align}
In what follows, several properties of the subgrid dissipation are discussed, resulting into requirements for subgrid-scale models.
By orthogonality of the terms of \cref{eq:generalnonlinearmodel}, we have
\begin{align}
\label{eq:generalnonlinearmodelsubgriddissipation}
D_{\tau^{mod}} & = -\tr{ \tau^{mod} \LESAvg{S} } = -\alpha_1' I_1,
\end{align}
so that only one model coefficient has to be tuned to parametrize dissipative effects in turbulent flows.

\subsubsection{Vreman's model requirements}
\label{sec:modconstraintsSGSTKEVreman}
Vreman~\cite{vreman04} requires that the modeled production of subgrid-scale kinetic energy vanishes for flows for which the actual production is known to be zero.
Preferably, also the opposite is true, i.e., if for a certain flow it is known that there is energy transport to subgrid scales, the model should cause the same behavior.
We can summarize Vreman's model requirements in the following form:
\begin{alignat}{2}
\label{eq:modreqVremana}
\mbox{P2a: } D_{\tau^{mod}} & = 0    & \mbox{ when } D_{\tau} & = 0, \\
\label{eq:modreqVremanb}
\mbox{P2b: } D_{\tau^{mod}} & \neq 0 & \mbox{ when } D_{\tau} & \neq 0.
\end{alignat}
To study the behavior of the subgrid dissipation, Vreman developed a classification of flows based on the number and position of zero elements in the (unfiltered) velocity gradient tensor. 
A total of 320 flow types can be distinguished, corresponding to all incompressible velocity gradients having zero to nine vanishing elements. 
Nonzero elements are left unspecified. 
Vreman shows that, for general filters, there are only thirteen flow types for which $D_{\tau}$ always vanishes. He calls such flow types locally laminar. Assuming an isotropic filter, we add another three flow classes. It can be shown that the subgrid dissipation is not generally zero for any of the other 304 flow classes.

On the basis of this classification of flows, we can analyze what Vreman calls the flow algebra of functionals of the velocity gradient, that is, the set of all flow types for which these functionals vanish.
A summary of the size of the flow algebra of different quantities is provided in \cref{tab:modconstraintsVreman}. 
Results can be compared to the desired outcome, as listed next to $D_{\tau}$. 
Not a single quantity was found that shows exactly the same behavior as the true subgrid dissipation.
This contrasts with Vreman's findings, which is due to the fact that we assume the use of a filter that conforms to the symmetry properties of the Navier-Stokes equations and, thus, is isotropic.
Eddy-viscosity-type models that are constructed using quantities that have a smaller flow algebra than the actual subgrid dissipation can be expected to be too dissipative. On the other hand, a model based on a quantity that is zero more often than $D_{\tau}$, can be expected to be underly dissipative.
Note that according to the current arguments locally two-dimensional flows exist that have a nonzero production of subgrid-scale kinetic energy.

\subsubsection{Nicoud \emph{et al.} model requirements}
\label{sec:modconstraintsSGSTKENicoud}
A different line of reasoning is followed by Nicoud \emph{et al.}~\cite{nicoudetal11}, who argue that a two-dimensional flow cannot be maintained if energy is transported to subgrid scales. 
They therefore see it as a desirable property that the modeled subgrid dissipation vanishes for all two-dimensional flows (P3a) and for the pure axisymmetric strain (P3b).
The quantities $I_4$, $I_6$ and $I_5 - \frac{ 1 }{ 2 } I_1 I_2$ satisfy both these properties.

It should be noted that properties P3a and P3b cannot be reconciled with Vreman's model requirement \hyperref[eq:modreqVremanb]{P2b}. Apparently, the physical reasoning employed by Nicoud \emph{et al.}~\cite{nicoudetal11} is not compatible with the mathematical properties of the turbulent stress tensor that were discovered by Vreman~\cite{vreman04}. For comparison we will, however, not exclude any requirements in what follows.

\subsubsection{Consistency with the second law of thermodynamics}
\label{sec:modconstraintsSGSTKE2ndlaw}

In turbulent flows, energy can be transported from large to small scales (forward scatter) and vice versa (backscatter). 
The second law of thermodynamics requires that the net transport of energy is of the former type~\cite{razafindralandyetal07}. 
Considering the fact that both subgrid and viscous dissipation play a role in large-eddy simulation, we need,
\begin{align}
\label{eq:modreq2ndlaw}
\begin{split}
\mbox{P4: } D_{\tau^{mod}} \geq 2 \nu D_{\LESAvg{S}} = - 2 \nu I_1.
\end{split}
\end{align}
In this context it is useful to note that $I_1$, $-I_2$, $-I_5$ and $I_5 - \frac{ 1 }{ 2 } I_1 I_2$ are all nonnegative quantities.

\section{Analysis of existing subgrid-scale models}
\label{sec:modanalysis}

Before looking at examples of subgrid-scale models that satisfy the constraints discussed in the previous section, let us first analyze the properties of existing models. 
A commonly used class of `eddy viscosity' models arises when it is assumed that small-scale turbulent motions effectively cause diffusion of the larger scales,
\begin{align}
\label{eq:modeddy2}
\begin{split}
\tau_e^{mod} - \frac{ 1 }{ 3 } \tr{\tau_e^{mod}} I  & = \alpha_1' \LESAvg{S} = -2 \nu_e \LESAvg{S}.
\end{split}
\end{align}
Turbulence is described as an essentially dissipative process by these models.
Examples of eddy viscosity models, written using the tensor invariants of \cref{eq:tensorinvariants}, are~\cite{smagorinsky63,nicoudducros99,vreman04,verstappenetal14,triasetal15}
\begin{alignat}{2}
	\label{eq:modSmag}
	& \mbox{Smagorinsky: }  &  \nu_e^{S} & = ( C_{S} \LESAvg{\delta} )^2 \sqrt{ 2 I_1 }, \\
	\label{eq:modWALE}
	& \mbox{WALE: }  &  \nu_e^{W} & = ( C_{W} \LESAvg{\delta} )^2 \frac{ J^{3/2} }{ I_1^{5/2} + J^{5/4} },
	\mbox{ where } J = \frac{ 1 }{ 6 } ( I_1 + I_2 )^2 + 2(I_5 - \frac{ 1 }{ 2 } I_1 I_2), \\
	\label{eq:modVreman}
	& \mbox{Vreman: }  &  \nu_e^{V} & = ( C_{V} \LESAvg{\delta} )^2 \sqrt{ \frac{ Q_G }{ P_G } }, \\
	\label{eq:modQR}
	& \mbox{QR: }  &  \nu_e^{QR} & = ( C_{QR} \LESAvg{\delta} )^2 \frac{ \max\{0, I_3\} }{ I_1 }, \\
	\label{eq:modS3PQR}
	& \mbox{S3PQR: }  &  \nu_e^{S3} & = ( C_{S3} \LESAvg{\delta} )^2 P_G^p Q_G^{-( p + 1 )} R_G^{ ( p + 5/2 )/3 },
\end{alignat}
where $\LESAvg{\delta}$ represents the large-eddy simulation filter length. In \cref{eq:modVreman,eq:modS3PQR}, the quantities
\begin{align}
\label{eq:modgradPQR}
\begin{alignedat}{3}
P_G & = I_1 - I_2, \qquad & 
Q_G & = \frac{ 1 }{ 4 } ( I_1 + I_2 )^2 + 4 (I_5 - \frac{ 1 }{ 2 } I_1 I_2 ), \qquad & 
R_G & = \frac{ 1 }{ 9 } ( I_3 + 3 I_4 )^2,
\end{alignedat}
\end{align}
are the tensor invariants of the gradient model~\cite{clarketal79},
\begin{align}
\label{eq:modgrad}
\begin{split}
\tau_G^{mod} & = C_G \LESAvg{\delta}^2 ( \LESAvg{S}^2 - \LESAvg{\Omega}^2 - (\LESAvg{S}\LESAvg{\Omega} - \LESAvg{\Omega}\LESAvg{S})).
\end{split}
\end{align}
Note that in contrast to the eddy viscosity models of \cref{eq:modSmag,eq:modWALE,eq:modVreman,eq:modQR}, this nonlinear model can account for backscatter and nondissipative processes. 
A different nonlinear model is the explicit algebraic subgrid-scale stress model (EASSM)~\cite{marstorpetal09}, which can be written
\begin{align}
\label{eq:modEASSM}
\begin{split}
\tau_E^{mod} & = \frac{ 4 }{ 3 } C_{E} \LESAvg{\delta}^2 I_1 I
- \frac{ C_{E}^{1.75} \LESAvg{\delta}^2 \gamma_1 I_1 }{ C_{E}^{1.5} \gamma_2 I_1 - \gamma_3 I_2 } \sqrt{I_1} \LESAvg{S}
- \frac{ C_{E} \LESAvg{\delta}^2 \gamma_4 I_1 }{ C_{E}^{1.5} \gamma_2 I_1 - \gamma_3 I_2 } ( \LESAvg{S}\LESAvg{\Omega} - \LESAvg{\Omega}\LESAvg{S} ),
\end{split}
\end{align}
for numerical constants $\gamma_i$.
\cref{tab:modanalysis} provides a summary of the behavior of the above subgrid-scale models with respect to the model requirements discussed in \cref{sec:modconstraints}. A detailed analysis of results is omitted, but note that existing models do not necessarily satisfy all the desired properties.

\begin{table}
	\centering
	\caption{
		Summary of the size of the flow algebra of the true subgrid dissipation $D_{\tau}$, \cref{eq:subgriddissipation}, and of several quantities based on the tensor invariants of \cref{eq:tensorinvariants}. $P_G$, $Q_G$ and $R_G$, \cref{eq:modgradPQR}, are the three principal invariants of the gradient model, \cref{eq:modgrad}. $D_G$ is its subgrid dissipation. 
		$Q_n$ represents the set of flow types for which the velocity gradient contains $n$ zero elements. 
		The total number of flows (3D), and the number of locally two-dimensional (2D) flows, \cref{eq:2Dflowchar}, are listed for reference. 
		Results provided here differ slightly from those of Vreman~\cite{vreman04}, because we assume the use of an isotropic filter.
	}
	\label{tab:modconstraintsVreman}
	\begin{tabular} {lccccccccccc}
		\toprule
		                                & $Q_0$ & $Q_1$ & $Q_2$ & $Q_3$ & $Q_4$ & $Q_5$ & $Q_6$ & $Q_7$ & $Q_8$ & $Q_9$ & $Q_{0-9}$ \\
		\midrule
		3D flows                        & 1     & 9     & 33    & 66    & 81    & 66    & 39    & 18    & 6     & 1     & 320       \\
		2D flows                        &       &       &       &       &       & 3     & 6     & 12    & 6     & 1     & 28        \\
		$D_\tau$                        &       &       &       &       &       &       &       & 9     & 6     & 1     & 16        \\
		\midrule
		$I_1$                           &       &       &       &       &       &       &       &       &       & 1     & 1         \\
		$I_2$                           &       &       &       &       &       &       & 1     & 3     &       & 1     & 5         \\
		$I_3$                           &       &       &       &       &       & 6     & 18    & 18    & 6     & 1     & 49        \\
		$I_4$                           &       &       &       &       &       & 6     & 19    & 18    & 6     & 1     & 50        \\
		$I_5$                           &       &       &       &       &       &       & 1     & 3     &       & 1     & 5         \\
		$I_6$                           &       &       &       &       & 3     & 15    & 19    & 12    & 6     & 1     & 56        \\
		\midrule
		$I_1 + I_2$                     &       &       &       &       &       &       & 8     & 12    & 6     & 1     & 27        \\
		$I_5 - \frac{ 1 }{ 2 } I_1 I_2$ &       &       &       &       &       & 3     & 7     & 12    & 6     & 1     & 29        \\
		$P_G$                           &       &       &       &       &       &       &       &       &       & 1     & 1         \\
		$Q_G$                           &       &       &       &       &       &       &       & 6     & 6     & 1     & 13        \\
		$R_G$                           &       &       &       & 6     & 30    & 48    & 36    & 18    & 6     & 1     & 145       \\
		$D_G = I_3 - I_4$               &       &       &       &       &       & 6     & 20    & 18    & 6     & 1     & 51        \\
		\bottomrule
	\end{tabular}
\end{table}

\begin{table}
	\centering
	\caption{	
		Summary of properties of several subgrid-scale models. 
		The properties considered are 
		\hyperref[eq:nssymmtimetrans]{S1}, \hyperref[eq:nssymmprestrans]{S2}, \hyperref[eq:nssymmgenGaltrans]{S3}, \hyperref[eq:nssymmorthtrans]{S4}:~time, pressure, generalized Galilean, and rotation and reflection invariance;
		\hyperref[eq:nssymmscaltrans]{S5}:~scaling invariance;
		\hyperref[eq:nssymm2DMFI]{S6}:~two-dimensional material frame-indifference;
		\hyperref[eq:eusymmtimerev]{S7}:~time reversal invariance;
		\hyperref[sec:modconstraintsnearwallscal]{P1}:~the proper near-wall scaling behavior, \cref{sec:modconstraintsnearwallscal};
		\hyperref[eq:modreqVremana]{P2a}:~zero subgrid dissipation for laminar flow types;
		\hyperref[eq:modreqVremanb]{P2b}:~possibly nonzero subgrid dissipation for nonlaminar flow types;
		\hyperref[sec:modconstraintsSGSTKENicoud]{P3a}:~zero subgrid dissipation for 2D flows, \cref{sec:modconstraintsSGSTKENicoud};
		\hyperref[sec:modconstraintsSGSTKENicoud]{P3b}:~zero subgrid dissipation for the pure axisymmetric strain;
		\hyperref[eq:modreq2ndlaw]{P4}:~consistency with the second law of thermodynamics.
		A blank space indicates a dependence on the model constants.
		*The dynamic procedure may restore these properties.
		**Depending on the value of the model parameter, $p$. Not a single parameter can ensure that all properties are satisfied.
	}
	\label{tab:modanalysis}
	\begin{tabular} {lccccccccc}
		\toprule
		     & \textbf{Smag.} & \textbf{WALE} & \textbf{Vreman} & \textbf{QR} & \textbf{S3PQR} & \textbf{Grad.} & \textbf{EASSM} & \textbf{Ex. 1} & \textbf{Ex. 2} \\
		\namecref{eq:modSmag} & \labelcref{eq:modSmag} & \labelcref{eq:modWALE} & \labelcref{eq:modVreman} & \labelcref{eq:modQR} & \labelcref{eq:modS3PQR} & \labelcref{eq:modgrad} & \labelcref{eq:modEASSM} & \labelcref{eq:modexI1I3} & \labelcref{eq:modexI5I1I2} \\
		\midrule
		S1-4 & Yes            & Yes           & Yes             & Yes         & Yes            & Yes            & Yes            & Yes            & Yes            \\
		S5*  & No             & No            & No              & No          & No             & No             & No             & No             & No             \\
		S6   & Yes            & No            & No              & Yes         & Yes**          & No             & No             & Yes            & Yes            \\
		S7*  & No             & No            & No              & No          & Yes**          & Yes            & No             & Yes            & No             \\
		P1   & No             & Yes           & No              & No          & Yes            & No             & No             & Yes            & Yes            \\
		P2a  & No             & No            & No              & Yes         & Yes**          & Yes            & No             & Yes            & Yes            \\
		P2b  & Yes            & Yes           & Yes             & No          & No             & No             & Yes            & No             & No             \\
		P3a  & No             & No            & No              & Yes         & Yes**          & Yes            & No             & Yes            & Yes            \\
		P3b  & No             & No            & No              & No          & No             & No             & No             & No             & Yes            \\
		P4   & Yes            & Yes           & Yes             & Yes         & Yes**          &                & Yes            &                & Yes            \\
		\bottomrule
	\end{tabular}
\end{table}

\section{Examples of physically-consistent subgrid-scale models}
\label{sec:modexamples}

In \cref{sec:modconstraints} we have seen how model requirements lead to constraints on the coefficients of the class of nonlinear models, \cref{eq:generalnonlinearmodel}. 
We will now strive to find the functional dependence of the model coefficients on the tensor invariants of \cref{eq:tensorinvariants} that satisfies these constraints.
Here it is to be noted that not all constraints can be satisfied simultaneously. 
As mentioned before, the model requirements of Nicoud \emph{et al.}, \cref{sec:modconstraintsSGSTKENicoud}, are incompatible with those of Vreman (in particular \hyperref[eq:modreqVremanb]{P2b}). 
This is because the former requirements are based on physical arguments that apparently do not match with the mathematical properties of the turbulent stresses on which the latter are based.
Furthermore, no 2D MFI~(\hyperref[eq:nssymm2DMFI]{S6}) quantities were found that satisfy both of Vreman's requirements, \cref{sec:modconstraintsSGSTKEVreman}. 
This may point to a limitation of assumption \cref{eq:assumption}.
It turns out that when compatible constraints are combined the model coefficients are not fully determined.
We thus obtain a class of what we will call physically-consistent subgrid-scale models. As far as the symmetries of the Navier-Stokes equations~(\hyperref[eq:nssymmtimetrans]{S1}-\hyperref[eq:eusymmtimerev]{S7}) and consistency with the second law of thermodynamics~(\hyperref[eq:modreq2ndlaw]{P4}) are concerned, this class of models extends the result of~\cite{razafindralandyetal07} by inclusion of the rate-of-rotation tensor.

The simplest physically-consistent models that exhibit the proper near-wall scaling behavior~(\hyperref[sec:modconstraintsnearwallscal]{P1}) have coefficients that depend only on the invariants of the rate-of-strain tensor, $I_1$ and $I_3$, e.g.,
\begin{align}
\label{eq:modexI1I3}
\begin{split}
\tau^{mod} & = c_0 \LESAvg{\delta}^2 \frac{ I_3^4 }{ I_1^5 } I + c_1 \LESAvg{\delta}^2 \frac{ I_3^3 }{ I_1^4 } \LESAvg{S} + c_2 \LESAvg{\delta}^2 \frac{ I_3^4 }{ I_1^6 } T_2' + c_3 \LESAvg{\delta}^2 \frac{ I_3^2 }{ I_1^3 } T_3' + c_4 \LESAvg{\delta}^2 \frac{ I_3^4 }{ I_1^6 } ( \LESAvg{S}\LESAvg{\Omega} - \LESAvg{\Omega}\LESAvg{S} ) + c_5 \LESAvg{\delta}^2 \frac{ I_3 }{ I_1^2 } T_5'.
\end{split}
\end{align}
This model satisfies all the symmetries of the Navier-Stokes equations, apart from scale invariance~(\hyperref[eq:nssymmscaltrans]{S5}). 
Consistency with the second law~(\hyperref[eq:modreq2ndlaw]{P4}) can be guaranteed by a proper choice of model constant $c_1$.

In view of the requirements of Nicoud \emph{et al.}, \cref{sec:modconstraintsSGSTKENicoud}, a possibly attractive model of eddy viscosity type, \cref{eq:modeddy2}, is based on $I_5 - \frac{ 1 }{ 2 } I_1 I_2$,
\begin{align}
\label{eq:modexI5I1I2}
\begin{split}
\nu_e & = ( C \LESAvg{\delta} )^2 \sqrt{ I_1 } (\frac{ 1 }{ 2 } - \frac{ I_5 }{ I_1 I_2 })^{3/2}.
\end{split}
\end{align}
It has the desired near-wall scaling behavior~(\hyperref[sec:modconstraintsnearwallscal]{P1}) and it vanishes only in flows that are locally two-dimensional, or in a state of pure shear or pure rotation.

For comparison, the properties of these example models are summarized in \cref{tab:modanalysis}.

\section{Conclusions and discussion}
\label{sec:concldisc}

We studied the construction of subgrid-scale models for large-eddy simulation of incompressible turbulent flows, aiming to preserve important mathematical and physical properties of the Navier-Stokes equations and the turbulent stresses.
As a starting point, we assumed an isotropic constitutive relation for the turbulent stresses in terms of the local filtered velocity gradient. From this we obtained a class of nonlinear models, consisting of independent terms, that allow for a description of both dissipative and nondissipative processes. 
We then discussed the symmetries of the Navier-Stokes equations, and the near-wall scaling and dissipation behavior of the turbulent stresses, and derived from them constraints on the coefficients of the class of nonlinear models.
Looking to satisfy these constraints by making the model coefficients depend explicitly on the local velocity gradient, we analyzed the behavior of different functions of this quantity. 
From this we obtained a class of physically-consistent models, of which we provided simple examples.
For comparison several existing subgrid-scale models were analyzed and it was noted that they do not all exhibit the desired properties.

A few remarks relating to our results are in place.
First of all, we derived a class of subgrid-scale models based on an assumption of locality of the turbulent stresses. It may turn out that nonlocal effects are important and additional transport equations have to be considered to obtain accurate results.
Secondly, recall that scale invariance is not guaranteed for models that are based on the local velocity gradient alone. This warrants a detailed analysis of the symmetry preservation properties of the dynamic procedure~\cite{germanoetal91} and other techniques that provide estimates of the turbulent kinetic energy and dissipation rate, like the integral length scale approximation~\cite{piomellietal15}.
These techniques may also lift the problem that, apart from a bound for the amount of dissipation coming from the second law of thermodynamics, the current model requirements do not provide information about the exact values of model constants.
Finally note that even some very successful models do not satisfy all of the discussed requirements. Of course adaptations to these models can now be suggested, but perhaps this observation can also motivate an assessment of the practical importance of each of the model requirements.

In future research we will look to expand the set of model requirements, for instance by consideration of realizability conditions for the turbulent stresses~\cite{vremanetal94}. 
Also, we will further study the behavior of the subgrid dissipation for exact solutions of the Navier-Stokes equations, and we propose to ensure that the modeled subgrid force vanishes for such solutions.
Finally, numerical tests of the proposed physically-consistent models are planned.

\section{Acknowledgments}
The authors thankfully acknowledge Professor Martin Oberlack for stimulating discussions during several stages of this project. Theodore Drivas and Perry Johnson are thankfully acknowledged for their valuable comments and criticisms on a preliminary version of this paper. 
Portions of this research have been presented at the 15th European Turbulence Conference, August 25-28th, 2015, Delft, The Netherlands.
This work is part of the Free Competition in Physical Sciences, which is financed by the Netherlands Organisation for Scientific Research (NWO). MHS gratefully acknowledges support from the Institute for Pure and Applied Mathematics (Los Angeles) for visits to the ``Mathematics of Turbulence'' program during the fall of 2014.

\begingroup
\setlength\bibitemsep{0pt}	
\printbibliography
\endgroup

\end{document}